\documentclass[11pt]{article}
\usepackage{amssymb,amsmath,amsfonts}
\usepackage{graphicx}
\usepackage{graphics}
\usepackage{eepic,epsfig}

\begin{document}

\title{Vacuum polarization by a flat boundary \\
in cosmic string spacetime}
\author{E. R. Bezerra de Mello$^{1}$\thanks{%
E-mail: emello@fisica.ufpb.br}\, and A. A. Saharian$^{1,2}$\thanks{%
E-mail: saharian@ysu.am} \\
\\
\textit{$^1$Departamento de F\'{\i}sica-CCEN, Universidade Federal da Para%
\'{\i}ba}\\
\textit{58.059-970, Caixa Postal 5.008, Jo\~{a}o Pessoa, PB, Brazil}\vspace{%
0.3cm}\\
\textit{$^2$Department of Physics, Yerevan State University,}\\
\textit{1 Alex Manoogian Street, 0025 Yerevan, Armenia}}
\maketitle

\begin{abstract}
In this paper we analyze the vacuum expectation values of the field squared
and the energy-momentum tensor associated to a massive scalar field in a
higher dimensional cosmic string spacetime, obeying Dirichlet or Neumann
boundary conditions on the surface orthogonal to the string. In order to
develop this analysis the corresponding Green function is obtained. The
Green function is given by the sum of two expressions: the first one
corresponds to the standard Green function in the boundary-free cosmic
string spacetime and the second contribution is induced by the boundary. The
boundary induced parts have opposite signs for Dirichlet and Neumann
scalars. Because the analysis of vacuum polarization effects in the
boundary-free cosmic string spacetime have been developed in the literature,
here we are mainly interested in the calculations of the effects induced by
the boundary. In this way closed expressions for the corresponding
expectation values are provided, as well as their asymptotic behavior in
different limiting regions is investigated. We show that the non-trivial
topology due to the cosmic string enhances the boundary induced vacuum
polarization effects for both field squared and the energy-momentum tensor,
compared to the case of a boundary in Minkowski spacetime. The presence of
the cosmic string induces non-zero stress along the direction normal to the
boundary. The corresponding vacuum force acting on the boundary is
investigated.
\end{abstract}

\bigskip

PACS numbers: 98.80.Cq, 11.10.Gh, 11.27.+d

\bigskip

\section{Introduction}

Cosmic strings are topologically stable gravitational defects which may have
been created in the early Universe after Planck time by a vacuum phase
transition \cite{Kibble,V-S}. The gravitational field produced by a cosmic
string may be approximated by a planar angle deficit in the two-dimensional
sub-space. The simplest theoretical model which describes a straight and
infinitely long cosmic string is given by a Dirac-delta type distribution
for the energy-momentum tensor along the linear defect. Also this object can
be described by classical field theory, coupling the energy-momentum tensor
associated with the Maxwell-Higgs system investigated by Nielsen and Olesen
in \cite{N-O} with the Einstein equations \cite{Garfinkle,Linet}. Though the
recent observational data on the cosmic microwave background have ruled out
cosmic strings as the primary source for primordial density perturbation,
they are still candidate for the generation of a number of interesting
physical effects such as gamma ray bursts \cite{Berezinski}, gravitational
waves \cite{Damour} and high energy cosmic rays \cite{Bhattacharjee}.
Moreover, in the framework of brane inflation \cite{Sarangi}-\cite{Dvali},
cosmic strings have attracted renewed interest, partly because a variant of
their formation mechanism is proposed.

Although the geometry of the spacetime produced by an idealized cosmic
string is locally flat, the planar angle deficit provides nonzero vacuum
expectation values (VEVs) for different physical observables. In this
context, the VEVs of the energy-momentum tensor have been calculated for
scalar and fermionic fields in \cite{scalar}-\cite{scalar4} and \cite{ferm}-%
\cite{ferm3}, respectively. Another type of vacuum polarization takes place
when boundaries are present. In this sense, by imposing boundary conditions
on quantum fields, additional shifts in the VEVs of physical quantities,
such as the energy density and stresses, take place. This is the well-known
Casimir effect (for a review see \cite{Cas}). The analysis of the Casimir
effect in the idealized cosmic string spacetime have been developed for
scalar \cite{Mello}, vector \cite{Brev95,Mello1} and fermionic fields \cite%
{Aram1,Beze10}, obeying boundary conditions on cylindrical surfaces.\footnote{%
Also vacuum polarization effects induced by a composite
topological defect have been analyzed in \cite{Mello3}.} The
Casimir force for massless scalar fields subject to Dirichlet and
Neumann boundary conditions in the setting of the conical piston
has been recently investigated in \cite{Fucc11}. Continuing along
this line of investigation, in this paper we shall analyze the
contribution on the vacuum polarization effects in a
higher-dimensional cosmic string spacetime induced by a scalar
field obeying Dirichlet or Neumann boundary conditions on a
surface orthogonal to the string. In addition to be a new
perspective related with the Casimir effect, this present
investigation may be relevant in the analysis of vacuum
polarization effects induced by a brane in anti-de Sitter
spacetime.

The paper is organized as follows. In section \ref{sec2} we provide the
general expression for the scalar Green function in a higher-dimensional
cosmic string spacetime admitting that the field obeys Dirichlet or Neumann
conditions on a boundary orthogonal to the string. We shall see that this
Green function is expressed in terms of two distinct contributions. The
first one is the standard Green function for a massive scalar field in a
boundary-free cosmic string spacetime, and the second contribution is due to
the boundary condition obeyed by the field operator. The first contribution
to the Green function is divergent at the coincidence limit; as to the
second one, it is finite in this limit for points away from the boundary.
Moreover, for specific values of the parameter associated with the planar
angle deficit, the complete Green function can be expressed in a closed form
in terms of a finite sum of the Macdonald functions. Because the analysis of
the VEVs of physical quantities induced by the cosmic string have been
developed in literature by many authors, in sections \ref{sec3} and \ref%
{sec4} we shall calculate the contributions to the VEVs of the field squared
and the energy-momentum tensor induced by the boundary. We shall see that
near the boundary and for points outside the string these contributions
become more relevant than the contributions due to the cosmic string itself.
In section \ref{conc}, we summarize the most important results obtained. In
this paper we shall use the units $\hbar =c=1$.

\section{Green function}

\label{sec2}

\subsection{Geometry of the problem and the heat kernel}

We consider a massive scalar quantum field propagating in a $D$-dimensional
cosmic string spacetime. By using the generalized cylindrical coordinates
with the cosmic string on the subspace defined by $r=0$, being $r\geqslant 0$
the radial polar coordinate, the corresponding metric tensor is defined by
the line element below:
\begin{equation}
ds^{2}=g_{ik}dx^{i}dx^{k}=-dt^{2}+dr^{2}+\alpha ^{2}r^{2}d\varphi
^{2}+dz^{2}+\sum_{l=4}^{D-1}(dx^{l})^{2}\ .  \label{cs1}
\end{equation}%
The coordinate system reads: $x^{i}=(t,r,\varphi ,z,x^{l})$, with $\varphi
\in \lbrack 0,\ 2\pi ]$, and $t,\ z,\ x^{l}\in (-\infty ,\ \infty )$. The
parameter $\alpha $, smaller than unity, codifies the presence of the
string. In a 4-dimensional spacetime, this parameter is related to the
linear mass density of the string by $\alpha =1-4G\mu $, with $G$ being the
Newton gravitational constant. In this analysis we shall admit the presence
of extra coordinates, $x^{l}$, defined in a Euclidean $(D-4)$-dimensional
subspace.

For a scalar field propagating in an arbitrary curved spacetime the field
equation has the form:
\begin{equation}
\left( \Box -m^{2}-\xi R\right) \phi (x)=0\ ,  \label{eq1}
\end{equation}%
with $\Box $ denoting the covariant d'Alembertian and $R$ is the scalar
curvature. In (\ref{eq1}) we have introduced an arbitrary curvature coupling
$\xi $. The minimal coupling corresponds to $\xi =0$ and for the conformal
one $\xi =\xi _{c}=(D-2)/[4(D-1)]$. We shall assume that the field obeys
Dirichlet boundary condition on the hypersurface orthogonal to the string
and located at $z=0$:
\begin{equation}
\phi (x)=0\ ,\ z=0\ .  \label{bc}
\end{equation}

The Green function associated with a massive scalar field in a curved
spacetime obeys the second order differential equation
\begin{equation}
\left( \Box -m^{2}-\xi R\right) G(x,x^{\prime })=-\delta ^{D}(x,x^{\prime
})=-\frac{\delta ^{D}(x-x^{\prime })}{\sqrt{-g}}\ ,  \label{Green1}
\end{equation}%
where $\delta ^{D}(x,x^{\prime })$ represents the bidensity Dirac
distribution. This Green function can be obtained within the framework of
the Schwinger-DeWitt formalism as follows:
\begin{equation}
G(x,x^{\prime })=\int_{0}^{\infty }ds\ {\mathcal{K}}(x,x^{\prime };s)\ ,
\label{heat}
\end{equation}%
where the heat kernel, ${\mathcal{K}}(x,x^{\prime };s)$, can be expressed in
terms of a complete set of normalized eigenfunctions of the operator defined
in (\ref{Green1}) as follows:
\begin{equation}
{\mathcal{K}}(x,x^{\prime };s)=\sum_{\sigma }\Phi _{\sigma }(x)\Phi _{\sigma
}^{\ast }(x^{\prime })e^{-s\sigma ^{2}}\ ,  \label{heat1}
\end{equation}%
with $\sigma ^{2}$ being the corresponding positively defined eigenvalue.

Writing
\begin{equation}
\left( \Box -m^{2}-\xi R\right) \Phi _{\sigma }(x)=-\sigma ^{2}\Phi _{\sigma
}(x)\ ,
\end{equation}%
in the spacetime defined by the line element (\ref{cs1}), a complete set of
normalized solutions of the above equation, compatible with the boundary
condition (\ref{bc}), can be specified in terms of a set of quantum numbers $%
(\omega ,\ q,\ n,\ k_{z},\ k_{l})$, where $n=0,\ \pm 1,\ \pm 2,\ ...\ $, $%
(\omega ,\ k_{l})\in \ (-\infty ,\ \infty )$ and $(q,\ k_{z})\geqslant 0$.
These functions are given by:
\begin{equation}
\Phi _{\sigma }(x)=2\sqrt{\frac{q}{\alpha }}\frac{e^{i(n\varphi +\mathbf{k}%
\cdot \mathbf{x}-\omega t)}}{(2\pi )^{(D-1)/2}}J_{|n|/\alpha }(qr)\sin
(k_{z}z)\ ,  \label{sol}
\end{equation}%
being $J_{\nu }(x)$ the Bessel function and $\mathbf{x}=(x^{4},\ldots
,x^{D-1})$. The corresponding positively defined eigenvalue is given below:
\begin{equation}
\sigma ^{2}=\omega ^{2}+q^{2}+k_{z}^{2}+\mathbf{k}^{2}+m^{2}\ .
\end{equation}

The heat kernel can be given in terms of the above eigenfunctions according
to (\ref{heat1}). After performing the integrals with the help of \cite{Grad}%
, we obtain:
\begin{equation}
{\mathcal{K}}(x,x^{\prime };s)=\frac{2e^{-\frac{\Delta \rho ^{2}}{4s}-sm^{2}}%
}{\alpha (4s\pi )^{D/2}}\sinh \left( \frac{zz^{\prime }}{2s}\right)
\sum_{n=-\infty }^{+\infty }e^{in\Delta \varphi }I_{|n|/\alpha }\left( \frac{%
rr^{\prime }}{2s}\right) \ ,  \label{heat2}
\end{equation}%
where $I_{\nu }(x)$ is the modified Bessel function and
\begin{equation}
\Delta \rho ^{2}=-\Delta t^{2}+r^{2}+r^{\prime }{}^{2}+\Delta \mathbf{x}%
^{2}+z^{2}+z^{\prime }{}^{2}\ ,
\end{equation}%
with $\Delta \varphi =\varphi -\varphi ^{\prime }$, $\Delta t=t-t^{\prime }$%
, $\Delta \mathbf{x=x-x}^{\prime }$.

In general, it is not possible to provide a closed expression for the Green
function by integrating over the variable $s$ the heat kernel function (\ref%
{heat2}), according to (\ref{heat}). However, for massless fields and for
specific values of the parameter $\alpha $, the corresponding Green
functions can be expressed in terms of a finite sum of the associated
Legendre functions and the Macdonald ones, respectively. These two different
situations will be analyzed separately in the following subsections.

The case of a scalar field with Neumann boundary condition,
$\partial _{z}\phi =0$ at $z=0$, can be considered in a similar
way. The corresponding eigenfunctions have the form (\ref{sol})
with the replacement $\sin (k_{z}z)\rightarrow \cos (k_{z}z)$. The
expression for the heat kernel is obtained from (\ref{heat2}) with
the replacement $\sinh (zz^{\prime }/(2s))\rightarrow \cosh
(zz^{\prime }/(2s))$.

\subsection{Special case}

\label{special}

The analysis of vacuum polarization effects associated with a quantum scalar
field in a cosmic string spacetime have been developed by many authors for
the case where the parameter $\alpha $ is equal to the inverse of an integer
number $p$, i.e., when $\alpha =1/p$ (see \cite{scalar1}-\cite{scalar3}). In
this case the corresponding Green function can be expressed in terms of $p$
images of the Minkowski spacetime function. Recently the image method was
also used in \cite{Mello} to obtain a closed expressions for massive scalar
Green functions in a higher-dimensional cosmic string space-time obeying
Robin boundary condition on a cylindrical surface coaxial with the string.
Here, in this subsection, we shall consider this specific situation, i.e., $%
\alpha $ being the inverse of an integer number, to obtain the Green
function in a closed form for the physical situation under consideration.
For this case, the expression of the heat kernel can be further simplified
with the help of the formula \cite{Pru,Jean}:
\begin{equation}
\sum_{n=-\infty }^{+\infty }e^{in\Delta \varphi }I_{|n|p}(rr^{\prime }/2s)=%
\frac{1}{p}\sum_{k=0}^{p-1}e^{\frac{rr^{\prime }}{2s}\cos (\frac{\Delta
\varphi }{p}+\frac{2\pi k}{p})}\ .  \label{Kaa}
\end{equation}%
The corresponding heat kernel reads,
\begin{equation}
{\mathcal{K}}(x,x^{\prime };s)=\frac{2e^{-sm^{2}}}{(4s\pi )^{D/2}}\sinh
\left( \frac{zz^{\prime }}{2s}\right) \sum_{k=0}^{p-1}e^{-\frac{{\mathcal{V}}%
_{k}}{4s}}\ ,
\end{equation}%
where
\begin{equation}
{\mathcal{V}}_{k}=-\Delta t^{2}+\Delta \mathbf{x}^{2}+z^{2}+z^{\prime
}{}^{2}+r^{2}+r^{\prime }{}^{2}-2rr^{\prime }\cos \left( \Delta \varphi
/p+2\pi k/p\right) \ .
\end{equation}

Finally, substituting the above function into (\ref{heat}), with the help of
\cite{Grad}, we get,
\begin{equation}
G(x,x^{\prime })=\frac{m^{D-2}}{(2\pi )^{D/2}}\sum_{k=0}^{p-1}\left[
f_{D/2-1}\left( m{\mathcal{V}}_{k(-)}\right) -f_{D/2-1}\left( m{\mathcal{V}}%
_{k(+)}\right) \right] \ ,  \label{g-special}
\end{equation}%
where
\begin{equation}
{\mathcal{V}}_{k(\mp )}=\left[ -\Delta t^{2}+\Delta \mathbf{x}^{2}+(z\mp
z^{\prime })^{2}+r^{2}+r^{\prime }{}^{2}-2rr^{\prime }\cos \left( \Delta
\varphi /p+2\pi k/p\right) \right] ^{1/2}\ .
\end{equation}%
In (\ref{g-special}) and in what follows we use the notation%
\begin{equation}
f_{\nu }(x)=K_{\nu }(x)/x^{\nu },  \label{fnu}
\end{equation}%
being $K_{\nu }(x)$ the Macdonald function. The expression (\ref{g-special})
is further simplified in the case of a massless field. By using the
asymptotic of the Macdonald function for small values of the argument \cite%
{Abra}, one finds
\begin{equation}
G(x,x^{\prime })=\frac{\Gamma (D/2-1)}{4\pi ^{D/2}}\sum_{k=0}^{p-1}\left[ {%
\mathcal{V}}_{k(-)}^{2-D}-{\mathcal{V}}_{k(+)}^{2-D}\right] \ .
\label{g-specialm0}
\end{equation}

We can see that the Green function (\ref{g-special}) vanishes for $z$ or $%
z^{\prime }$ being equal to zero. It can be presented as the sum of two
different contributions as shown below:
\begin{equation}
G(x,x^{\prime })=G_{\text{cs}}(x,x^{\prime })+G_{\text{b}}(x,x^{\prime })\ .
\label{GFdec}
\end{equation}%
The first term in the right-hand side coincides with the Green function for
a massive scalar field in the absence of the boundary. It is divergent at
the coincidence limit and the divergence comes from the $k=0$ term. As to
the second contribution, $G_{\text{b}}(x,x^{\prime })$, it is a consequence
of the boundary condition imposed on the field. This contribution is finite
at the coincidence limit for points outside the boundary.

The renormalized Green function, used for the evaluation of finite and well
defined VEVs, is given by subtracting from the corresponding Green function
the Hadamard one. Because the cosmic string spacetime is locally flat, the
Hadamard function coincides with the Green function in the Minkowski
spacetime. For the above Green function the renormalization procedure can be
implemented explicitly by discarding the $k=0$ component of the function $G_{%
\text{cs}}(x,x^{\prime })$.

The formulae for the Green function in the case of Neumann boundary
condition are obtained from (\ref{g-special}) and (\ref{g-specialm0})
changing the sign of the terms with ${\mathcal{V}}_{k(+)}$. Consequently,
the boundary induced parts in the Green function, $G_{\text{b}}(x,x^{\prime
})$, for Dirichlet and Neumann scalars differ by the sign. This is the case
also for the corresponding parts in the VEVs of the field squared and the
energy-momentum tensor discussed below in section \ref{sec3} and \ref{sec4}.

\subsection{General case}

For general case where $p=1/\alpha $ is not an integer number, the Green
function can be expressed in an integral form by substituting the heat
kernel (\ref{heat2}) into (\ref{heat}). After some intermediate steps the
Green function is presented in the form (\ref{GFdec}), where the
boundary-free and boundary induced parts are given by the expressions
\begin{eqnarray}
G_{\text{cs}}(x,x^{\prime }) &=&\frac{p}{(4\pi )^{D/2}}\sum_{n=-\infty
}^{+\infty }e^{in\Delta \varphi }\int_{0}^{\infty }dw\ w^{D/2-2}e^{-\frac{{%
\mathcal{V}_{(-)}}}{4}w-\frac{m^{2}}{w}}\ I_{|n|p}\left( rr^{\prime
}w/2\right) ,  \notag \\
G_{\text{b}}(x,x^{\prime }) &=&-\frac{p}{(4\pi )^{D/2}}\sum_{n=-\infty
}^{+\infty }e^{in\Delta \varphi }\int_{0}^{\infty }dw\ w^{D/2-2}e^{-\frac{%
\mathcal{V}_{(+)}}{4}w-\frac{m^{2}}{w}}\ I_{|n|p}\left( rr^{\prime
}w/2\right) \ ,  \label{GFcsb}
\end{eqnarray}%
where
\begin{equation}
{\mathcal{V}_{(\mp )}}=-\Delta t^{2}+\Delta \mathbf{x}^{2}+(z\mp z^{\prime
})^{2}+r^{2}+r^{\prime }{}^{2}\ .  \label{Vpm}
\end{equation}%
We can also verify that the Green function vanishes for $z$ or $z^{\prime }$
equal to zero. Here we can analyze the behavior of the both contributions in
the coincidence limit. In this limit, ${\mathcal{V}_{(-)}}=2r^{2}$, so due
to the exponential behavior of the modified Bessel function for large
arguments, the first integral diverges for large values of the variable $w$;
as to the second contribution, ${\mathcal{V}_{(+)}}=4z^{2}+2r^{2}$,
consequently for $z\neq 0$ the integrand in the expression for $G_{\text{b}%
}(x,x^{\prime })$ goes to zero exponentially for large values of the
integration variable.

We can also provide a closed expression for the Green function in the limit
of a massless field. With the help of \cite{Grad} we find:
\begin{eqnarray}
G(x,x^{\prime }) &=&\frac{e^{-i(D-3)\pi /2}}{(2\pi )^{(D+1)/2}}\frac{p}{%
(rr^{\prime })^{D/2-1}}\sum_{n=-\infty }^{+\infty }e^{in\Delta \varphi }
\notag \\
&&\times \left[ \frac{Q_{|n|p-1/2}^{(D-3)/2}(\cosh u_{(-)})}{(\sinh
u_{(-)})^{(D-3)/2}}-\frac{Q_{|n|p-1/2}^{(D-3)/2}(\cosh u_{(+)})}{(\sinh
u_{(+)})^{(D-3)/2}}\right] \ ,  \label{Leg1}
\end{eqnarray}%
being $Q_{\nu }^{\mu }(x)$ the associated Legendre function and
\begin{equation}
\cosh u_{(\mp )}=\frac{{\mathcal{V}_{(\mp )}}}{2rr^{\prime }}\geqslant 1\ .
\label{chu}
\end{equation}%
Finally, we have to say that the renormalized Green function can be obtained
by subtracting from the Green function the corresponding function in the
Minkowski spacetime, which is given by $G_{\text{cs}}(x,x^{\prime })$ taking
$p=1$.

\section{Vacuum expectation value of the field squared}

\label{sec3} This and the following sections will be devoted to the
calculations of vacuum polarizations effects induced by the boundary. Two
main calculations will be performed. The evaluation of the VEV of the field
squared, in the first place, followed by the evaluation of the VEV of the
energy-momentum tensor.

The VEV of the field squared is formally given by evaluating the Green
function at the coincidence limit. In this analysis the complete Green
function is given by the sum of the Green function in the cosmic string
spacetime in the absence of the boundary plus the boundary induced part. In
this way we may write,
\begin{equation}
\langle \phi ^{2}\rangle =\langle \phi ^{2}\rangle _{\text{cs}}+\langle \phi
^{2}\rangle _{\text{b}}\ .  \label{phi2sum}
\end{equation}%
However, because the singular behavior of $G_{\text{cs}}(x,x^{\prime })$ at
the coincidence limit, the renormalization procedure is needed for the first
contribution of the above expression. The second contribution is finite at
the coincidence limit for points outside the hypersurface $z=0$. Because the
VEV of the field squared in the cosmic string spacetime has been analyzed by
many authors, here we are mainly interested in the analysis of the quantum
effects induced by the boundary.

According to the previous section, we shall analyze the VEV of the field
squared induced by the boundary for $p$ being an integer number in the first
part, and for the general case in the second one.

\subsection{Special case}

Being $p$ an integer number, the VEV of the field squared induced by the
boundary is given simply by taking the coincidence limit of $G_{\text{b}%
}(x^{\prime },x)$ given in (\ref{g-special}). The result is presented by the
expression
\begin{equation}
\langle \phi ^{2}\rangle _{\text{b}}=-\frac{m^{D-2}}{(2\pi )^{D/2}}%
\sum_{k=0}^{p-1}f_{D/2-1}(2m\sqrt{z^{2}+r^{2}s_{k}^{2}})\ ,
\label{P2-special}
\end{equation}%
where
\begin{equation}
s_{k}=\sin (\pi k/p)\ .  \label{sk}
\end{equation}%
As it is seen, the boundary induced part in the VEV is negative. In (\ref%
{P2-special}), the part with the term $k=0$ is the corresponding VEV in the
Minkowski spacetime and we can write:
\begin{equation}
\langle \phi ^{2}\rangle _{\text{b}}=\langle \phi ^{2}\rangle _{\text{b}%
}^{(p=1)}+\langle \phi ^{2}\rangle _{\text{b}}^{\text{C}}\ ,  \label{phi2BC}
\end{equation}%
where the second term on the right-hand side is the part of the VEV induced
by the non-trivial topology of the cosmic string spacetime. Note that for
the renormalized pure topological part one has the expression (see \cite%
{Mello})
\begin{equation}
\langle \phi ^{2}\rangle _{\text{cs}}=\frac{m^{D-2}}{(2\pi )^{D/2}}%
\sum_{k=1}^{p-1}f_{D/2-1}\left( 2mrs_{k}\right) .  \label{phi2cs}
\end{equation}%
The latter is always positive. It is of interest to note that the
topological part in the total VEV, $\langle \phi ^{2}\rangle _{\text{cs}%
}+\langle \phi ^{2}\rangle _{\text{b}}^{\text{C}}$, vanishes on the boundary.

\begin{figure}[tbph]
\begin{center}
\epsfig{figure=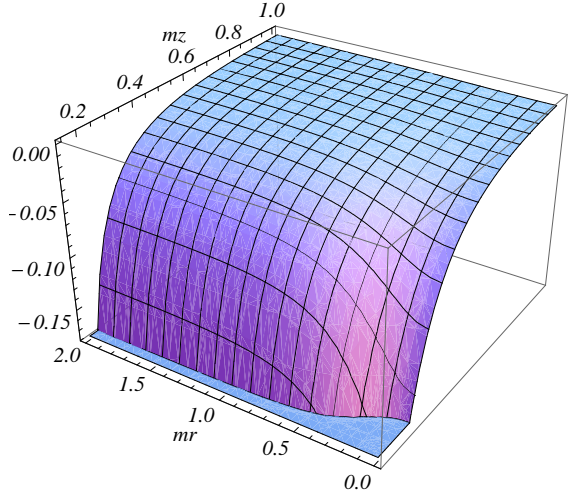, width=7.5cm, height=6.5cm}
\end{center}
\caption{The boundary induced part in the VEV\ of the field squared, $%
\langle \protect\phi ^{2}\rangle _{\text{b}}/m^{2}$, for a 4-dimensional
cosmic string spacetime with $p=3$, as a function of $mr$ and $mz$.}
\label{fig1}
\end{figure}

For $z\neq 0$ and for points far from the string, $r\gg |z|$, the dominant
contribution to the boundary induced part (\ref{P2-special}) comes from the $%
k=0$ term and to the leading order we get%
\begin{equation}
\langle \phi ^{2}\rangle _{\text{b}}\approx \langle \phi ^{2}\rangle _{\text{%
b}}^{(p=1)}=-\frac{m^{D-2}}{(2\pi )^{D/2}}f_{D/2-1}\left( 2m|z|\right) .
\label{phi20}
\end{equation}%
Note that for $z\neq 0$ and for points on the string axis, $r=0$, one has $%
\langle \phi ^{2}\rangle _{\text{b},r=0}=p\langle \phi ^{2}\rangle _{\text{b}%
}^{(p=1)}$. In the limit $|z|\gg r$ and $m|z|\gg 1$, the leading order term
provides an exponentially suppressed behavior below,
\begin{equation}
\langle \phi ^{2}\rangle _{\text{b}}\approx -\frac{p\ m^{(D-3)/2}e^{-2m|z|}}{%
2(4\pi )^{(D-1)/2}|z|^{(D-1)/2}}\ .  \label{phi2bsp}
\end{equation}%
For a massless field, from (\ref{P2-special}) we obtain
\begin{equation}
\langle \phi ^{2}\rangle _{\text{b}}=-\frac{\Gamma (D/2-1)}{(4\pi )^{D/2}}%
\sum_{k=0}^{p-1}(z^{2}+r^{2}s_{k}^{2})^{1-D/2}\ .  \label{phi2bsp0}
\end{equation}%
In this case, at large distances from the boundary, $|z|\gg r$, the boundary
induced VEV decays as power-law: $\langle \phi ^{2}\rangle _{\text{b}%
}\propto r^{2-D}$. In figure \ref{fig1} we exhibit the behavior of (\ref%
{P2-special}) as a function of the dimensionless variables $mr$ and $mz$ for
$D=4$, $p=3$.

As we have already explained and it is seen from the graph, for fixed $z$
the boundary induced contribution is finite for $z\neq 0$ and $r=0$. On the
other hand, for fixed non-vanishing radial coordinate, near the boundary it
is dominated by the $k=0$ term (see (\ref{phi20})). When $z$ goes to
infinity the VEV is exponentially suppressed.

\begin{figure}[tbph]
\begin{center}
\epsfig{figure=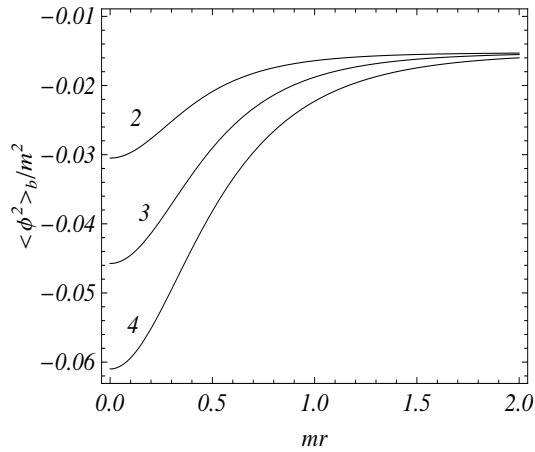, width=7.cm, height=6.cm}
\end{center}
\caption{This graph provides the behavior of $\langle \protect\phi %
^{2}\rangle _{\text{b}}/m^{2}$ in a 4-dimensional spacetime as a function of
$mr$ for $mz=0.5$ and for several values of the parameter $p$ (numbers near
the curves). }
\label{fig2}
\end{figure}

In figure \ref{fig2} we exhibit the behavior of (\ref{P2-special}) in the
case of a $D=4$ cosmic string as a function of the dimensionless variable $%
mr $ for three distinct values of $p=2,\ 3,\ 4$ and for $mz=0.5$. We can see
that the effects induced by the string become more relevant for larger
values of $p$.

\subsection{General case}

In the case of $p$ being not an integer number, the VEV of the field squared
induced by the boundary is obtained by taking the coincidence limit of $G_{%
\text{b}}(x^{\prime },x)$ given in (\ref{GFcsb}). After a change of the
integration variable it can be written as:
\begin{equation}
\langle \phi ^{2}\rangle _{\text{b}}=-\frac{pr^{2-D}}{(2\pi )^{D/2}}%
\int_{0}^{\infty }\ dy\ y^{D/2-2}e^{-\left( 2z^{2}/r^{2}+1\right)
y-m^{2}r^{2}/(2y)}\ S_{p}(y)\ ,  \label{P2-general}
\end{equation}%
with
\begin{equation}
S_{p}(y)=\sideset{}{'}{\sum}_{n=0}^{\infty }I_{np}(y)\ ,  \label{Sp}
\end{equation}%
where the prime means that the term with $n=0$ should be halved. As it is
seen from (\ref{P2-general}), the boundary induced VEV is negative for
general case of $p$. Although being not possible to provide a closed
expression for the above VEV, some limiting cases can be obtained. Due to
the exponential decay, for $|z|\gg r$ the dominant contribution comes from
the region near the lower limit of the integration. In this region we may
approximate $S_{p}(y)\approx 1/2$ and with the help of \cite{Grad} we obtain
$\langle \phi ^{2}\rangle _{\text{b}}\approx p\langle \phi ^{2}\rangle _{%
\text{b}}^{(p=1)}$. We can see that for $p>1$ the presence of the cosmic
string increases the above results when compared with the corresponding ones
in the absence of it.

We can find an integral representation for the VEV of the field squared by
using the formula
\begin{equation}
S_{p}(y)=\frac{1}{p}\sideset{}{'}{\sum}_{k=0}^{p_{0}}e^{y\cos (2\pi k/p)}-%
\frac{\sin (p\pi )}{2\pi }\int_{0}^{\infty }dx\frac{e^{-y\cosh x}}{\cosh
(px)-\cos (p\pi )},  \label{SumForm}
\end{equation}%
where $p_{0}$ is an integer number defined by $2p_{0}<p<2p_{0}+2$. For even
values of $p$ the corresponding formula is obtained from (\ref{SumForm}) by
taking the limit. In this limit the second term on the right-hand side
becomes $e^{-y}/(2p)$. The formula (\ref{SumForm}) is obtained as a special
case of the more general formula derived in \cite{Beze10}.

Substituting (\ref{SumForm}) into (\ref{P2-general}), the integration over $%
y $ is done explicitly in terms of the Macdonald function and one finds
\begin{eqnarray}
\langle \phi ^{2}\rangle _{\text{b}} &=&-\frac{2m^{D-2}}{(2\pi )^{D/2}}\ %
\bigg[\sideset{}{'}{\sum}_{k=0}^{p_{0}}f_{D/2-1}(2m\sqrt{z^{2}+r^{2}s_{k}^{2}%
})  \notag \\
&&-\frac{p}{\pi }\sin (p\pi )\int_{0}^{\infty }dx\frac{f_{D/2-1}(2m\sqrt{%
z^{2}+r^{2}\cosh ^{2}x})}{\cosh (2px)-\cos (p\pi )}\bigg].  \label{phi2b}
\end{eqnarray}%
It can be seen that for integer values $p$ this result reduces to the
expression (\ref{P2-special}). On the string axis for the boundary induced
VEV one has%
\begin{equation}
\langle \phi ^{2}\rangle _{\text{b},r=0}=p\langle \phi ^{2}\rangle _{\text{b}%
}^{(p=1)}.  \label{phi2bAxis}
\end{equation}%
For points near the boundary, $|z|\ll m^{-1}$, $|z|\ll r$, the main
contribution to the boundary induced VEV comes from the term $k=0$ and to
the leading order we find%
\begin{equation}
\langle \phi ^{2}\rangle _{\text{b}}\approx -\frac{\Gamma (D/2-1)}{(4\pi
)^{D/2}|z|^{D-2}}.  \label{phi2Bnear}
\end{equation}%
This leading term does not depend on the angle deficit and it coincides with
the corresponding term for a boundary in Minkowski spacetime. At large
distances from the boundary, $|z|\gg m^{-1}$, the boundary induced part is
exponentially suppressed and the VEV is dominated by the pure topological
part $\langle \phi ^{2}\rangle _{\text{cs}}$.

For a massless field, by using the relation $f_{\nu }(y)\sim 2^{\nu
-1}\Gamma (\nu )y^{-2\nu }$, $y\rightarrow 0$, we find%
\begin{eqnarray}
\langle \phi ^{2}\rangle _{\text{b}} &=&-\frac{2\Gamma (D/2-1)}{(4\pi )^{D/2}%
}\ \bigg[\sideset{}{'}{\sum}_{k=0}^{p_{0}}(z^{2}+r^{2}s_{k}^{2})^{1-D/2}
\notag \\
&&-\frac{p}{\pi }\sin (p\pi )\int_{0}^{\infty }dx\frac{\left(
z^{2}+r^{2}\cosh ^{2}x\right) ^{1-D/2}}{\cosh (2px)-\cos (p\pi )}\bigg].
\label{phi2Bm0}
\end{eqnarray}%
In this case, at large distances from the boundary, $|z|\gg r$, the leading
term in the corresponding asymptotic expansion has the form
\begin{equation}
\langle \phi ^{2}\rangle _{\text{b}}\approx p\langle \phi ^{2}\rangle _{%
\text{b}}^{(p=1)}=-\frac{p\Gamma (D/2-1)}{(4\pi )^{D/2}|z|^{D-2}}.
\label{phi2bm0Large}
\end{equation}%
Recall that near the boundary we have the behavior given by (\ref{phi2Bnear}%
).

For a scalar field with Neumann boundary condition at $z=0$, the
corresponding formulae for the VEV of the field squared are obtained from
those given above by changing the sign of the boundary induced part.

\begin{figure}[tbph]
\begin{center}
\epsfig{figure=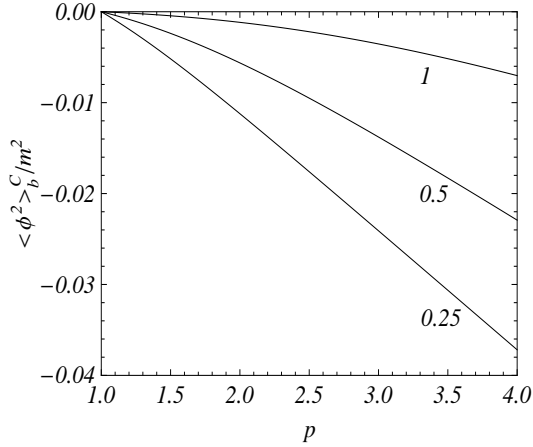, width=7cm, height=6.cm }
\end{center}
\caption{The topological part in the boundary induced VEV of the field
squared, $\langle \protect\phi ^{2}\rangle _{\text{b}}^{\text{C}}/m^{2}$, in
a 4-dimensional cosmic string spacetime as a function of $p$ for $mz=0.5$
and for several values of $mr$ (numbers near the curves). }
\label{fig3n}
\end{figure}

In figure \ref{fig3n} we plot $\langle \phi ^{2}\rangle _{\text{b}}^{\text{C}%
}/m^{2}$ as a function of the parameter $p$ for $mz=0.5$ and for several
values of $mr$ (numbers near the curves) in a 4-dimensional spacetime ($D=4$%
). Note that for the first term in the right-hand side of (\ref{phi2BC}),
for $mz=0.5$ one has $\langle \phi ^{2}\rangle _{\text{b}}^{(p=1)}\approx
-0.0152m^{2}$.

We can also analyze the topological part in the boundary induced VEV of the
field squared by adding and subtracting from (\ref{P2-general}) the part
corresponding to the Minkowski spacetime ($p=1$):
\begin{eqnarray}
\langle \phi ^{2}\rangle _{\text{b}} &=&-\frac{r^{2-D}}{(2\pi )^{D/2}}%
\int_{0}^{\infty }dyy^{N/2}e^{-\left( 2z^{2}/r^{2}+1\right)
y-m^{2}r^{2}/(2y)}\sideset{}{'}{\sum}_{n=0}^{\infty }I_{n}(y)  \notag \\
&-&\frac{r^{2-D}}{(2\pi )^{D/2}}\int_{0}^{\infty }dyy^{N/2}e^{-\left(
2z^{2}/r^{2}+1\right) y-m^{2}r^{2}/(2y)}\sideset{}{'}{\sum}_{n=0}^{\infty }%
\left[ pI_{pn}(y)-I_{n}(y)\right] .  \label{phi2rep2}
\end{eqnarray}%
The first contribution can be promptly obtained by noting that (see (\ref%
{SumForm})) $\sideset{}{'}{\sum}_{n=0}^{\infty }I_{n}(y)=e^{y}/2$. Of
course, the corresponding expression coincides with (\ref{phi20}) which is
independent of $r$ and diverges for $z=0$.

An alternative representation for the topological part can be obtained by
using the Abel-Plana formula (see, for instance, \cite{SahaRev}) for the
summation over $n$ in the second term on the right-hand side of (\ref%
{phi2rep2}):
\begin{equation}
\sideset{}{'}{\sum}_{n=0}^{\infty }F(n)=\int_{0}^{\infty }du\
F(u)+i\int_{0}^{\infty }du\ \frac{F(iu)-F(-iu)}{e^{2\pi u}-1}\ .
\end{equation}%
In our case, $F(n)=pI_{np}(y)$. Now we can see that in the evaluation of the
difference the terms coming from the first integral on the right-hand side
of the Abel-Plana formula cancel out and one obtains
\begin{equation}
\sideset{}{'}{\sum}_{n=0}^{\infty }\left[ pI_{np}(y)-I_{n}(y)\right] =\frac{2%
}{\pi }\int_{0}^{\infty }dv\ g(v,p)K_{iv}(y)\ ,  \label{ndif}
\end{equation}%
where we have introduced the notation
\begin{equation}
g(v,p)=\sinh (\pi v)\left( \frac{1}{e^{2\pi v/p}-1}-\frac{1}{e^{2\pi v}-1}%
\right) .  \label{gbu}
\end{equation}%
The respective contribution becomes:
\begin{eqnarray}
\langle \phi ^{2}\rangle _{\text{b}}^{\text{C}} &=&-\frac{4r^{2-D}}{(2\pi
)^{D/2+1}}\int_{0}^{\infty }dyy^{D/2-2}e^{-\left( 2z^{2}/r^{2}+1\right)
y-m^{2}r^{2}/(2y)}  \notag \\
&&\times \int_{0}^{\infty }dv\ g(v,p)K_{iv}(y)\ .  \label{phi2C}
\end{eqnarray}%
For $r\neq 0$ the above correction is finite at the boundary.

Taking $z=0$, for a massless field the integral over $y$ in (\ref{phi2C}) is
evaluated in terms of the gamma function and we obtain
\begin{equation}
\langle \phi ^{2}\rangle _{\text{b},z=0}^{\text{C}}=-\frac{8(4\pi
)^{-(D+1)/2}}{\Gamma ((D-1)/2)r^{D-2}}\int_{0}^{\infty }dv\ g(v,p)\left\vert
\Gamma \left( D/2-1-iv\right) \right\vert ^{2}.
\end{equation}%
For an even dimensional spacetime the modulus of the gamma function in this
formula is expressed in terms of elementary functions and the integral can
be explicitly evaluated. In particular, for $D=4$ one has $|\Gamma
(1-iv)|^{2}=\pi v/\sinh (\pi v)$ and the integral above becomes equal to $%
\pi (p^{2}-1)/24$. In this case we get:
\begin{equation}
\langle \phi ^{2}\rangle _{\text{b},z=0}^{\text{C}}=-\frac{p^{2}-1}{48\pi
^{2}r^{2}}\ .  \label{phi2z0}
\end{equation}%
This result is also obtained from (\ref{P2-special}), in the special case of
integer $p$. Taking $D=4$, for a massless fields we have:
\begin{equation}
\langle \phi ^{2}\rangle _{\text{b}}^{\text{C}}=-\frac{1}{16\pi ^{2}}%
\sum_{k=1}^{p-1}\frac{1}{z^{2}+r^{2}\sin ^{2}(\pi k/p)}\ .  \label{phi2Spz0}
\end{equation}%
On the boundary, $z=0$, the sum in this formula is explicitly evaluated, $%
\sum_{k=1}^{p-1}\sin ^{-2}(\pi k/p)=(p^{2}-1)/3$, and (\ref{phi2Spz0})
reduces to (\ref{phi2z0}).

\section{Energy-momentum tensor}

\label{sec4}

Following the same line of investigation, in this section we are interested
to calculate the contribution induced by the boundary on the VEV of the
energy-momentum tensor. Similar to the case of the field squared, the
energy-momentum tensor is presented in the decomposed form,
\begin{equation}
\langle T_{ik}\rangle =\langle T_{ik}\rangle _{\text{cs}}+\langle
T_{ik}\rangle _{\text{b}},  \label{EMTdec}
\end{equation}%
where $\langle T_{ik}\rangle _{\text{cs}}$ corresponds to the geometry of
the cosmic string without boundaries. In order to evaluate the boundary
induced part we shall use the following expression:
\begin{equation}
\langle T_{ik}\rangle _{\text{b}}=\lim_{x^{\prime }\rightarrow x}\partial
_{i^{\prime }}\partial _{k}G_{\text{b}}(x,x^{\prime })+\left[ \left( \xi -{1}%
/{4}\right) g_{ik}\Box -\xi \nabla _{i}\nabla _{k}-\xi R_{ik}\right] \langle
\phi ^{2}\rangle _{\text{b}}\ ,  \label{mvevEMT}
\end{equation}%
where for the spacetime under consideration the Ricci tensor, $R_{ik}$,
vanishes.

By using the expression (\ref{phi2b}), for the covariant d'Alembertian
appearing in (\ref{mvevEMT}) one finds
\begin{eqnarray}
&&\Box \langle \phi ^{2}\rangle _{\text{b}}=-\frac{8m^{D}}{(2\pi )^{D/2}}\ %
\bigg\{\sideset{}{'}{\sum}_{l=0}^{p_{0}}\left[
4m^{2}(z^{2}+s_{l}^{4}r^{2})f_{D/2+1}(w_{1})-(1+2s_{l}^{2})f_{D/2}(w_{1})%
\right]  \notag \\
&&\qquad -\frac{p\sin (p\pi )}{2\pi }\int_{0}^{\infty }dx\frac{\left(
w_{2}^{2}+m^{2}r^{2}\sinh ^{2}x\right) f_{D/2+1}(w_{2})-(2+\cosh
x)f_{D/2}(w_{2})}{\cosh (px)-\cos (p\pi )}\bigg\},  \label{Dalphi2}
\end{eqnarray}%
where%
\begin{eqnarray}
w_{1} &=&2m\sqrt{z^{2}+r^{2}s_{k}^{2}},  \notag \\
w_{2} &=&2m\sqrt{z^{2}+r^{2}\cosh ^{2}(x/2)}.  \label{w12}
\end{eqnarray}%
In the evaluation of the derivative term $\partial _{\phi ^{\prime
}}\partial _{\phi }G_{\text{b}}(x,x^{\prime })$ we need the expression for
the sum $\sum_{n=-\infty }^{+\infty }n^{2}I_{|n|p}\left( y\right) $. For the
summation of this series we use the relation%
\begin{equation}
\sum_{n=-\infty }^{+\infty }n^{2}I_{|n|p}\left( y\right) =2p^{-2}\left(
y^{2}\partial _{y}^{2}+y\partial _{y}-y^{2}\right) S_{p}(y),  \label{rel1}
\end{equation}%
and the formula (\ref{SumForm}). In the evaluation of the other derivative
terms we can put $\phi =\phi ^{\prime }$ before the differentiation. In this
case, by using the formula (\ref{SumForm}), for the boundary induced part in
the Green function one finds%
\begin{eqnarray}
G_{\text{b}}(x,x^{\prime })|_{\varphi ^{\prime }=\varphi } &=&-\frac{2m^{D-2}%
}{(2\pi )^{D/2}}\ \bigg[\sideset{}{'}{\sum}_{l=0}^{p_{0}}f_{D/2-1}(m\sqrt{%
\mathcal{V}_{(+)}-2rr^{\prime }\cos (2\pi l/p)})  \notag \\
&&-\frac{p\sin (p\pi )}{2\pi }\int_{0}^{\infty }dx\frac{f_{D/2-1}(m\sqrt{%
\mathcal{V}_{(+)}+2rr^{\prime }\cosh x})}{\cosh (px)-\cos (p\pi )}\bigg],
\label{GBphi}
\end{eqnarray}%
where $\mathcal{V}_{(+)}$ is defined by (\ref{Vpm}).

Combining the formulae given above, for the separate components of the
energy-momentum tensor we find the following VEVs:%
\begin{eqnarray}
&&\langle T_{i}^{l}\rangle _{\text{b}}=-\frac{2m^{D}}{(2\pi )^{D/2}}\ \bigg[%
\sideset{}{'}{\sum}_{k=0}^{p_{0}}F_{i}^{l}(2mr,2mz,s_{k})  \notag \\
&&\qquad -\frac{p}{\pi }\sin (p\pi )\int_{0}^{\infty }dx\frac{%
F_{i}^{l}(2mr,2mz,\cosh x)}{\cosh (2px)-\cos (p\pi )}\bigg].  \label{TilB}
\end{eqnarray}%
Here the functions are defined as%
\begin{eqnarray}
F_{0}^{0}(u,v,w) &=&\left( 4\xi -{1}\right) (w^{4}u^{2}+v^{2})f_{D/2+1}(y)+%
\left[ 1-\left( 4\xi -{1}\right) (1+2w^{2})\right] f_{D/2}(y),  \notag \\
F_{1}^{1}(u,v,w) &=&(4\xi -1)v^{2}f_{D/2+1}(y)+2[1-2\xi (1+w^{2})]f_{D/2}(y),
\notag \\
F_{2}^{2}(u,v,w) &=&\left[ 4\xi (w^{4}u^{2}+v^{2})-y^{2}\right]
f_{D/2+1}(y)+2[1-2\xi (1+w^{2})]f_{D/2}(y),  \label{Fil} \\
F_{3}^{3}(u,v,w) &=&(4\xi -1)w^{2}\left[ w^{2}u^{2}f_{D/2+1}(y)-2f_{D/2}(y)%
\right] ,  \notag \\
F_{3}^{1}(u,v,w) &=&-(4\xi -1)uvw^{2}f_{D/2+1}(y),  \notag
\end{eqnarray}%
the indices $0,1,2,3$ correspond to the coordinates $t,r,\varphi ,z$, and%
\begin{equation}
y=\sqrt{v^{2}+u^{2}w^{2}}.  \label{y}
\end{equation}%
For the components (no summation over $l$) $\langle T_{l}^{l}\rangle _{\text{%
b}}$, $l=4,\ldots ,D-1$, one has the relation $\langle T_{l}^{l}\rangle _{%
\text{b}}=\langle T_{0}^{0}\rangle _{\text{b}}$. Of course, the latter
relation is a direct consequence of the invariance of the problem with
respect to the boosts along the directions $x^{l}$, $l=4,\ldots ,D-1$. Due
to the presence of the boundary the boost invariance along the $z$-axis is
lost and, as a consequence, $\langle T_{3}^{3}\rangle _{\text{b}}\neq
\langle T_{0}^{0}\rangle _{\text{b}}$. As we see, the vacuum energy-momentum
tensor is non-diagonal. Note that the off-diagonal component vanishes for $%
z=0$ and $r=0$, separately. In addition, on the string axis, $r=0$, one has $%
\langle T_{2}^{2}\rangle _{\text{b}}=\langle T_{1}^{1}\rangle _{\text{b}}$.
For a scalar field with Neumann boundary condition at $z=0$, the
corresponding formula for the VEV of the energy-momentum tensor is obtained
from (\ref{TilB}) by changing the sign of the boundary induced part.

Due to the presence of the off-diagonal component, from the covariant
conservation condition, $\nabla _{\mu }\langle T_{\nu }^{\mu }\rangle =0$,
for the energy-momentum tensor two non-trivial differential equations
follow:
\begin{equation}
\partial _{r}(r\langle T_{1}^{1}\rangle _{\text{b}})+r\partial _{z}\langle
T_{1}^{3}\rangle _{\text{b}}=\langle T_{2}^{2}\rangle _{\text{b}}
\end{equation}%
and
\begin{equation}
\partial _{z}\langle T_{3}^{3}\rangle _{\text{b}}=-\frac{1}{r}\partial
_{r}(r\langle T_{3}^{1}\rangle _{\text{b}})\ .
\end{equation}%
It can be checked that these relations are obeyed by the VEV of the
energy-momentum tensor above.

The $k=0$ term in (\ref{TilB}) is the corresponding VEV for a flat boundary
in Minkowski spacetime. The corresponding expression takes the form (no
summation over $l$)%
\begin{equation}
\langle T_{l}^{l}\rangle _{\text{b}}^{(p=1)}=-\frac{m^{D}}{(2\pi )^{D/2}}\ %
\left[ \left( 4\xi -{1}\right) (2mz)^{2}f_{D/2+1}(2m|z|)+2\left( 1-2\xi
\right) f_{D/2}(2m|z|)\right] ,  \label{Tllp1}
\end{equation}%
for $l=0,1,2,4,\ldots ,D-1$, and $\langle T_{i}^{l}\rangle _{\text{b}%
}^{(p=1)}=0$ for the other components. Note that for a conformally coupled
massless field $\langle T_{l}^{l}\rangle _{\text{b}}^{(p=1)}=0$.

For points near the boundary, $|z|\ll m^{-1}$, $|z|\ll r$, one has%
\begin{equation}
\langle T_{i}^{l}\rangle _{\text{b}}\approx \langle T_{i}^{l}\rangle _{\text{%
b}}^{(p=1)}\approx -2(D-1)\ \left( \xi -\xi _{c}\right) \frac{\Gamma
(D/2)\delta _{i}^{l}}{(4\pi )^{D/2}|z|^{D}},  \label{Tttp1near}
\end{equation}%
for $l=0,1,2,4,\ldots ,D-1$, and these components diverge on the boundary.
For a conformally coupled field the leading term vanishes and the subleading
term should be taken:%
\begin{equation}
\langle T_{i}^{l}\rangle _{\text{b}}\approx \langle T_{i}^{l}\rangle _{\text{%
b}}^{(p=1)}\approx \ \frac{2m^{2}\Gamma (D/2)|z|^{2-D}\delta _{i}^{l}}{(4\pi
)^{D/2}\left( D-1\right) (D-2)}.  \label{Tttp1nearm0}
\end{equation}%
The corresponding energy density, $\langle T^{00}\rangle _{\text{b}%
}=-\langle T_{0}^{0}\rangle _{\text{b}}$, is negative for both minimally and
conformally coupled scalar fields. For the components $\langle
T_{3}^{3}\rangle _{\text{b}}$, $\langle T_{3}^{1}\rangle _{\text{b}}$, the $%
k=0$ term in (\ref{TilB}) vanishes and these components are finite for $%
r\neq 0$. Note that for $z\neq 0$ the boundary induced part in the VEVs is
finite on the string axis, $r=0$. By taking into account that the pure
topological part diverges on the string we conclude that the latter
dominates for points near the string.

At large distances from the string, $r\gg m^{-1}$, $r\gg |z|$, the dominant
contribution to the boundary induced VEV (\ref{TilB}) comes from the $k=0$
term and, to the leading order, the VEVs coincide with the corresponding
expressions for a boundary in Minkowski spacetime. In this limit the effects
due to the non-trivial topology of the cosmic string are exponentially
suppressed. At large distances from the boundary, $|z|\gg m^{-1}$, the
boundary induced VEVs in the components of the energy-momentum tensor are
suppressed by the factor $e^{-2m|z|}$.

In the case of a massless field, for the VEVs we find the expression%
\begin{eqnarray}
&&\langle T_{i}^{l}\rangle _{\text{b}}=-\frac{\Gamma (D/2)}{(4\pi )^{D/2}}\ %
\left[ \sideset{}{'}{\sum}_{k=0}^{p_{0}}\frac{F_{(0)i}^{l}(r,z,s_{k})}{%
(z^{2}+r^{2}s_{k}^{2})^{D/2+1}}\right.  \notag \\
&&\qquad \left. -\frac{p}{\pi }\sin (p\pi )\int_{0}^{\infty }dx\frac{%
F_{(0)i}^{l}(r,z,\cosh x)}{\cosh (2px)-\cos (p\pi )}\frac{1}{%
(z^{2}+r^{2}\cosh ^{2}x)^{D/2+1}}\right] .  \label{Tilm0}
\end{eqnarray}%
The functions for the separate components are defined by the expressions%
\begin{eqnarray}
F_{(0)0}^{0}(r,z,w) &=&\left\{ (4\xi -1)[(D-2)w^{2}-1]+1\right\} r^{2}w^{2}+
\left[ (4\xi -1)(D-1-2w^{2})+1\right] z^{2},  \notag \\
F_{(0)1}^{1}(r,z,w) &=&\left[ 2-4\xi (1+w^{2})\right] r^{2}w^{2}+\left[ 4\xi
\left( D-1-w^{2}\right) -D+2\right] z^{2},  \notag \\
F_{(0)2}^{2}(r,z,w) &=&\left[ 4\xi \left( (D-1)w^{2}-1\right) -D+2\right]
r^{2}w^{2}+\left[ 4\xi \left( D-1-w^{2}\right) -D+2\right] z^{2},
\label{Fil0} \\
F_{(0)3}^{3}(r,z,w) &=&(4\xi -1)w^{2}\left[ (D-2)w^{2}r^{2}-2z^{2}\right] ,
\notag \\
F_{(0)3}^{1}(r,z,w) &=&-D(4\xi -1)rzw^{2}.  \notag
\end{eqnarray}%
Now it can be checked that for a massless conformally coupled scalar field
the tensor $\langle T_{i}^{l}\rangle _{\text{b}}$ is traceless. Note that
for a massless field the VEV $\langle T_{i}^{l}\rangle _{\text{b}}^{(p=1)}$
is given by the right-hand side of (\ref{Tttp1near}) and vanishes for a
conformally coupled field. The expression on the right-hand side of (\ref%
{Tttp1near}) gives the leading term in the asymptotic expansion of $\langle
T_{i}^{l}\rangle _{\text{b}}$ at large distances from the string, $r\gg |z|$%
, for the components with $i=l=0,1,2,4,\ldots ,D-1$. In the same limit, the
other non-zero components decay as $\langle T_{3}^{3}\rangle _{\text{b}%
}\propto r^{-D}$ and $\langle T_{3}^{1}\rangle _{\text{b}}\propto r^{-D-1}$.
At large distances from the boundary, $|z|\gg r$, the boundary induced part
in the VEVs of the diagonal components behave as (no summation over $l$) $%
\langle T_{l}^{l}\rangle _{\text{b}}$ $\propto |z|^{-D}$, whereas for the
off-diagonal component one has $\langle T_{3}^{1}\rangle _{\text{b}}$ $%
\propto |z|^{-D-1}$.

In the case $p$ being an integer number, the general formula (\ref{TilB})
takes the form
\begin{equation}
\langle T_{i}^{l}\rangle _{\text{b}}=-\frac{m^{D}}{(2\pi )^{D/2}}\
\sum_{k=0}^{p-1}F_{i}^{l}(2mr,2mz,s_{k}),  \label{TilBSp}
\end{equation}%
with the functions $F_{i}^{l}(u,v,w)$ defined in (\ref{Fil}). Note that the
corresponding expression for the boundary-free part has the form \cite{Mello}
(no summation over $l$)%
\begin{equation}
\langle T_{l}^{l}\rangle _{\text{cs}}=\frac{m^{D}}{(2\pi )^{D/2}}%
\sum_{k=1}^{p-1}F_{\text{cs},l}(2mr,s_{k}),  \label{Tllcs}
\end{equation}%
where%
\begin{eqnarray}
F_{\text{cs},0}(u,w) &=&(1-4\xi )u^{2}w^{4}f_{D/2+1}(uw)-\left[ 1+2(1-4\xi
)w^{2}\right] f_{D/2}(uw),  \notag \\
F_{\text{cs},1}(u,w) &=&\left( 4\xi w^{2}-1\right) f_{D/2}(uw),  \label{Fcs}
\\
F_{\text{cs},2}(u,w) &=&\left( 4\xi w^{2}-1\right) \left[
f_{D/2}(uw)-u^{2}w^{2}f_{D/2+1}(uw)\right] ,  \notag
\end{eqnarray}%
and $F_{\text{cs},l}(u,w)=F_{\text{cs},0}(u,w)$ for $l=3,\ldots ,D-1$. For a
massless field the formulae (\ref{TilBSp}) and (\ref{Tllcs}) are simplified
to (no summation over $l$)%
\begin{eqnarray}
\langle T_{i}^{l}\rangle _{\text{cs}} &=&\frac{\Gamma (D/2)\delta _{i}^{l}}{%
2(4\pi )^{D/2}r^{D}}\sum_{k=1}^{p-1}\frac{F_{\text{cs},l}^{(0)}(s_{k})}{%
s_{k}^{D}},  \label{TilBSpm0p1} \\
\langle T_{i}^{l}\rangle _{\text{b}} &=&-\frac{\Gamma (D/2)}{2(4\pi )^{D/2}}%
\sum_{k=0}^{p-1}\frac{F_{(0)i}^{l}(r,z,s_{k})}{(z^{2}+r^{2}s_{k}^{2})^{D/2+1}%
},  \label{TilBSpm0}
\end{eqnarray}%
with the functions
\begin{eqnarray}
F_{\text{cs},l}^{(0)}(w) &=&\left( D-2\right) (1-4\xi
)w^{2}-1,\;l=0,3,\ldots ,D-1,  \notag \\
F_{\text{cs},1}^{(0)}(w) &=&\frac{F_{\text{cs},2}^{(0)}(w)}{1-D}=4\xi
w^{2}-1,  \label{Fcs0}
\end{eqnarray}
and the functions $F_{(0)i}^{l}(r,z,w)$ are given by (\ref{Fil0}). For a
conformally coupled massless scalar field the vacuum energy density $\langle
T^{00}\rangle $ is always positive.

\begin{figure}[tbph]
\begin{center}
\begin{tabular}{cc}
\epsfig{figure=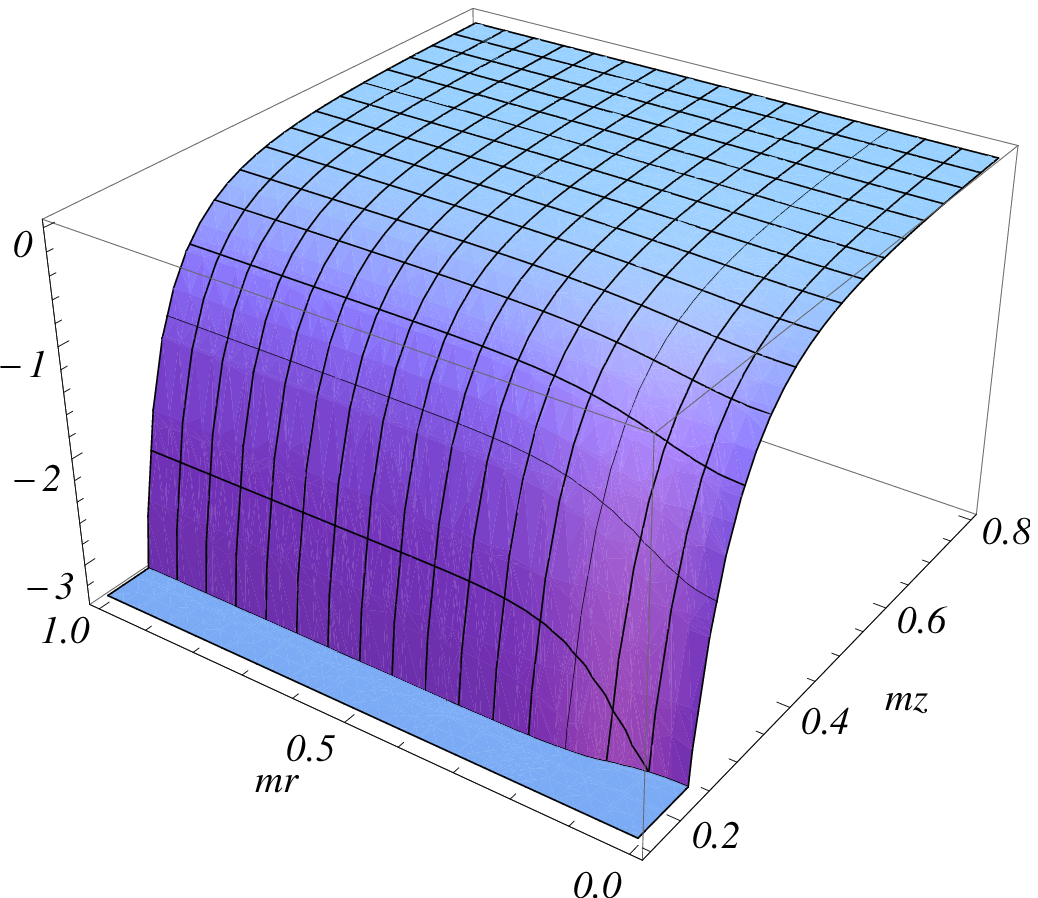, width=6.5cm, height=6.cm} & \quad %
\epsfig{figure=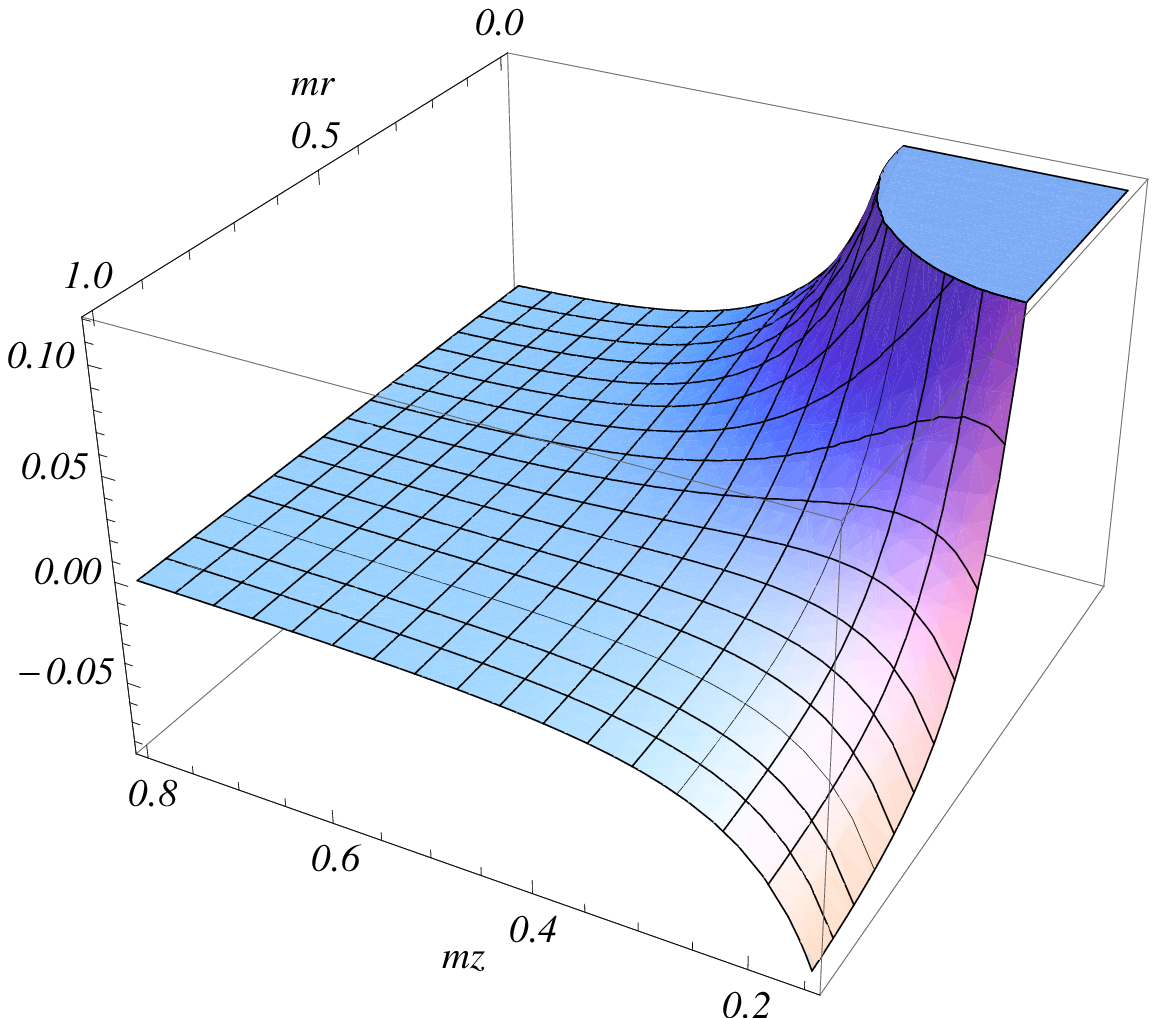, width=6.5cm, height=6.cm}%
\end{tabular}%
\end{center}
\caption{The boundary induced part in the VEV of the energy density, $%
\langle T^{00}\rangle _{\text{b}}/m{^{4}}$, for minimally (left plot) and
conformally (right plot) coupled scalar fields in a 4-dimensional spacetime
as a function of $mr$ and $mz$ for $p=3$.}
\label{fig4}
\end{figure}

In the left/right panel of figure \ref{fig4} the VEV of the energy density
induced by the boundary is exhibited for a minimally/conformally coupled
massive scalar field in a 4-dimensional spacetime, as a function of $mr$ and
$mz$ for $p=3$. We can see that the energy density crucially depends on the
curvature coupling parameter $\xi $.

The normal vacuum force acting on the boundary is finite for $r\neq 0$ and
is determined by the component $\langle T_{3}^{3}\rangle _{\text{b}}$
evaluated at $z=0$:%
\begin{eqnarray}
&&\langle T_{3}^{3}\rangle _{\text{b},z=0}=\frac{2m^{D}(1-4\xi )}{(2\pi
)^{D/2}}\ \bigg[\sum_{k=1}^{p_{0}}s_{k}^{2}F_{3}(2mrs_{k})  \notag \\
&&\qquad -\frac{p}{\pi }\sin (p\pi )\int_{0}^{\infty }dx\frac{F_{3}(2mr\cosh
x)\cosh ^{2}x}{\cosh (2px)-\cos (p\pi )}\bigg],  \label{Tzz0}
\end{eqnarray}%
where we have defined the function%
\begin{equation}
F_{3}(u)=u^{2}f_{D/2+1}(u)-2f_{D/2}(u).  \label{Fzu}
\end{equation}%
The vacuum effective pressure on the boundary is given by $P=\langle
T_{3}^{3}\rangle _{\text{b},z=0}$. Note that the dependence of the force on
the curvature coupling parameter appears in the form of the factor $(1-4\xi )
$. For a massless field, the expression (\ref{Tzz0}) reduces to
\begin{eqnarray}
\langle T_{3}^{3}\rangle _{\text{b},z=0} &=&(D-2)\frac{(1-4\xi )\Gamma (D/2)%
}{(4\pi )^{D/2}r^{D}}\ \bigg[\sum_{k=1}^{p_{0}}s_{k}^{2-D}  \notag \\
&&-\frac{p}{\pi }\sin (p\pi )\int_{0}^{\infty }dx\frac{\cosh ^{2-D}x}{\cosh
(2px)-\cos (p\pi )}\bigg].  \label{Tzz0m0}
\end{eqnarray}%
In particular, for $p$ being an integer number and for $D=4$ one has
\begin{equation}
\langle T_{3}^{3}\rangle _{\text{b},z=0}=\frac{1-4\xi }{48\pi ^{2}r^{4}}%
(p^{2}-1).  \label{Tzz0m0D4}
\end{equation}%
For the both minimally and conformally coupled scalar fields the
corresponding effective pressure is positive. In the left plot of figure \ref%
{fig5} we have presented $\langle T_{3}^{3}\rangle _{\text{b},z=0}/m^{D}$
for a minimally coupled scalar field in $D=4$ as a function of $mr$ for
separate values of the parameter $p$ (numbers near the curves). In the right
plot the same quantity is given as a function of $p$ for separate values of $%
mr$ (numbers near the curves).
\begin{figure}[tbph]
\begin{center}
\begin{tabular}{cc}
\epsfig{figure=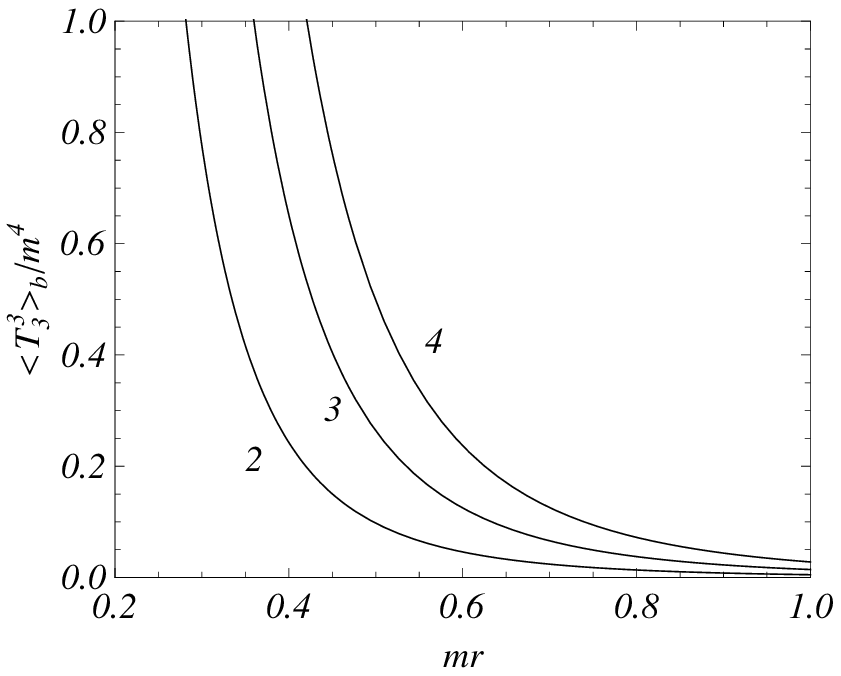, width=6.5cm, height=5.5cm} & \quad %
\epsfig{figure=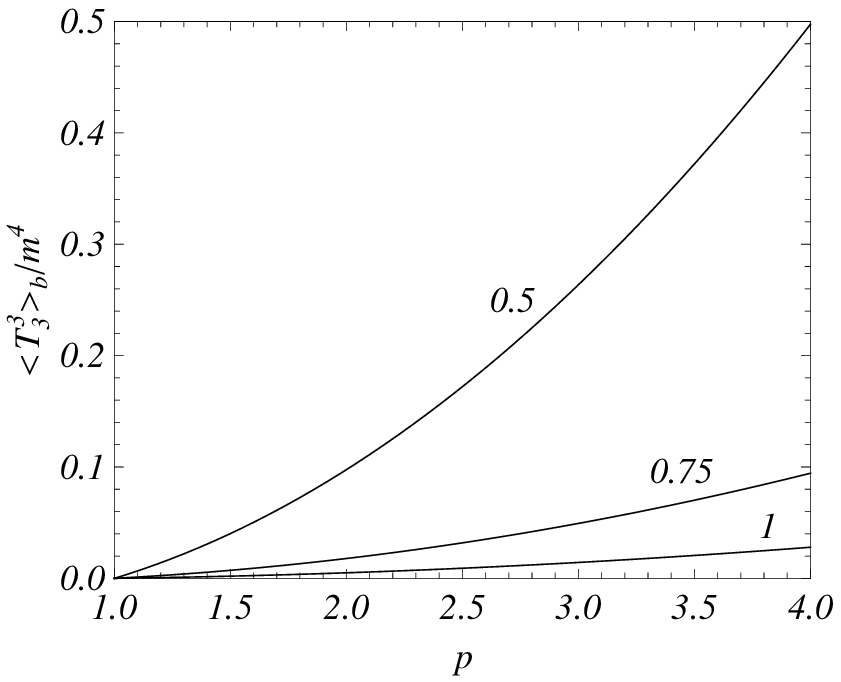, width=6.5cm, height=5.5cm}%
\end{tabular}%
\end{center}
\caption{The normal vacuum stress on the boundary, $\langle T_{3}^{3}\rangle
_{\text{b},z=0}/m^{D}$, for a minimally coupled scalar field in $D=4$ as a
function of $mr$ (left plot) for separate values of the parameter $p$
(numbers near the curves). In the right plot the same quantity is given as a
function of $p$ for separate values of $mr$ (numbers near the curves).}
\label{fig5}
\end{figure}

\section{Conclusion}

\label{conc} In this paper we have analyzed the effects induced by
a flat boundary on the VEVs of the field squared and the
energy-momentum tensor associated with a massive scalar field in a
higher-dimensional cosmic string spacetime. Although the analysis
of the VEVs associated with quantum fields in the cosmic string
spacetime taking into account the presence of boundaries, are, in
general, developed considering cylindrical surfaces coaxial with
the string, here we decided to adopt a different geometry by
considering a flat boundary surface orthogonal to the string. The
condition imposed on the field at the boundary is Dirichlet one.
For a scalar field with Neumann boundary condition, the
corresponding formulae for the VEVs of the field squared and the
energy-momentum tensor are obtained from those given above by
changing the sign of the boundary induced parts. In the presence
of the boundary, the VEVs obtained are given in terms of a sum of
two terms: the first one due to the cosmic string itself in the
absence of boundary and the second terms induced by the boundary.
Because the analysis of the VEVs associated with scalar fields in
a pure higher-dimensional cosmic string spacetime have been
developed in the literature, in this paper we were more interested
to investigate the contribution induced by the boundary. Two
distinct situations were considered: first one when the parameter
which codifies the presence of the string is the inverse of an
integer number, and in the second for general values of this
parameter. For the first case, the Green function is expressed in
terms of a finite sum of the Macdonald functions, while for the
second one, only an integral representation can be provided.

For integer values of the parameter $p$, the boundary induced part in the
VEV of the field squared is given by (\ref{P2-special}) and this part is
always negative. Note that the pure topological part is positive and is
given by the expression (\ref{phi2cs}). For general case of the parameter $p$%
, we have provided the integral representation (\ref{phi2b}). For points on
the string axis the boundary induced VEV in the field squared is related to
the corresponding quantity in Minkowski spacetime by simple formula (\ref%
{phi2bAxis}). For points near the boundary, the leading term in the
asymptotic expansion of the VEV does not depend on the angle deficit and it
coincides with corresponding term for a boundary in Minkowski spacetime. At
large distances from the boundary and for a massive field, the boundary
induced part is exponentially suppressed and the VEV of the field squared is
dominated by the pure topological part. For a massless field the boundary
induced VEV of the field squared is given by (\ref{phi2Bm0}) and at large
distances it decays as $|z|^{2-D}$.

Another important local characteristic of the vacuum state is the VEV of the
energy-momentum tensor. Similar to the field squared, the vacuum
energy-momentum tensor is decomposed into pure topological and boundary
induced parts. For the general case of the parameter $p$, the latter is
non-diagonal and is given by expression (\ref{TilB}). For a massless field
the corresponding formula is reduced to (\ref{Tilm0}). In the case $p$ being
an integer number, the general formulae take the forms (\ref{TilBSp}) and (%
\ref{TilBSpm0}) for massive and massless fields respectively. The
corresponding VEV in the boundary-free cosmic string geometry is given by (%
\ref{Tllcs}). For a conformally coupled massless scalar field the total
energy density is always positive. For points near the boundary, the VEVs of
the energy density and the diagonal stresses along the directions parallel
to the boundary diverge as $|z|^{-D}$ for a non-conformally coupled field.
For a conformally coupled field the leading terms vanish and these VEVs
behave as $|z|^{2-D}$. The VEVs of the normal stress, $\langle
T_{3}^{3}\rangle _{\text{b}}$, and of the off-diagonal component, $\langle
T_{3}^{1}\rangle _{\text{b}}$, are finite on the boundary for points outside
the string axis.

The normal vacuum force acting on the plate is determined by the component $%
\langle T_{3}^{3}\rangle _{\text{b}}$ evaluated on the boundary. This force
is determined by expressions (\ref{Tzz0}) and (\ref{Tzz0m0}) for massive and
massless fields respectively. In particular, for a massless field in
4-dimensional spacetime the corresponding effective pressure is given by (%
\ref{Tzz0m0D4}) and it is positive for both conformal and minimal couplings.
Note that for a flat boundary in Minkowski spacetime the normal stress
vanishes and the effective force in the geometry under consideration is
induced by the presence of the string. Numerical examples presented show
that the non-trivial topology due to the cosmic string enhances the vacuum
polarization effects for both field squared and the energy-momentum tensor
compared to the case of a boundary in Minkowski spacetime.

Before to finish this paper, we would like to say that the analysis
developed here may be relevant in the investigation of vacuum polarization
effects induced by a cosmic string in anti-de Sitter spacetime \cite%
{Cristine}. In fact for this geometry, adopting Poincar\'{e}
coordinate system, the coordinate along the string has a
semi-infinity range and a flat boundary orthogonal to the string
is present. It is also our future interest to analyze the vacuum
polarization effects associated with quantum fields in this
geometry.

\section*{Acknowledgment}

E.R.B.M. thanks Conselho Nacional de Desenvolvimento Cient\'{\i}fico e Tecnol%
\'{o}gico (CNPq) for partial financial support.


\begin{thebibliography}{99}
\bibitem{Kibble} T. W. Kibble, J. Phys. A \textbf{9}, 1387 (1976).

\bibitem{V-S} A. Vilenkin and E. P. S. Shellard, \textit{Cosmic Strings and
Other Topological Defects} (Cambridge University Press, Cambridge, England,
1994).

\bibitem{N-O} N. B. Nielsen and P. Olesen, Nucl. Phys. B\textbf{\ 61}, 45
(1973).

\bibitem{Garfinkle} D. Garfinkle, Phys. Rev. D \textbf{32}, 1323 (1985).

\bibitem{Linet} B. Linet, Phys. Lett. B \textbf{124}, 240 (1987).

\bibitem{Berezinski} V. Berezinski, B. Hnatyk and A. Vilenkin, Phys. Rev. D
\textbf{64}, 043004 (2001).

\bibitem{Damour} T. Damour and A. Vilenkin, Phys. Rev. Lett. \textbf{85},
3761 (2000).

\bibitem{Bhattacharjee} P. Bhattacharjee and G. Sigl, Phys. Rep. \textbf{327}%
, 109 (2000).

\bibitem{Sarangi} S. Sarangi and S. -H. Henry Tye, Phys. Lett. B \textbf{536}%
, 185 (2002).

\bibitem{Copeland} E. J. Copeland, R. C. Myers and J. Polchinski, J. High
Energy Phys. \textbf{06}, 013 (2004).

\bibitem{Dvali} G. Dvali and A. Vilenkin, J. Cosmol. Astropart. Phys.
\textbf{03}, 010 (2004).

\bibitem{scalar} B. Linet, Phys. Rev. D \textbf{35}, 536 (1987).

\bibitem{scalar1} A. G. Smith, in \textit{Symposium on the Formation and
Evolution of Cosmic String}, edited by G. W. Gibbons, S. W. Hawking and T.
Vachaspati (Cambridge University Press, Cambridge, England, 1989).

\bibitem{scalar2} P. C. Davies and V. Sahni, Class. Quantum Grav. \textbf{5}
1 (1987).

\bibitem{scalar3} T. Souradeep and V. Sahni, Phys. Rev. D \textbf{46}, 1616
(1992).

\bibitem{scalar4} M. E. X. Guimar\~aes and B. Linet, Class. Quantum Grav.
\textbf{10}, 1665 (1993).

\bibitem{ferm} V. P. Frolov and E. M. Serebriany, Phys. Rev. D \textbf{15},
3779 (1287).

\bibitem{ferm1} B. Linet, J. Math. Phys. \textbf{36}, 3694 (1995).

\bibitem{ferm2} E. S. Moreira Jnr., Nucl. Phys. B \textbf{451}, 365 (1995).

\bibitem{ferm3} V. B. Bezerra and N. R. Khusnutdinov, Class. Quantum Grav.
\textbf{23}, 3449 (2006).

\bibitem{Cas} E. Elizalde, S. D. Odintsov, A. Romeo, A. A. Bytsenko and S.
Zerbini, \textit{Zeta Regularization Techniques With Applications} (World
Scientific, Singapore, 1994); V. M. Mostepanenko and N. N. Trunov, \textit{%
The Casimir Effect and Its Applications} (Clarendon, Oxford, 1997); K.A.
Milton, \textit{The Casimir Effect: Physical Manifestation of Zero-Point
Energy} (World Scientific, Singapore, 2002); V. A. Parsegian, Van der Waals
Forces (Cambridge University Press, Cambridge, England, 2005); M. Bordag, G.
L. Klimchitskaya, U. Mohideen and V. M. Mostepanenko, \textit{Advances in
the Casimir Effect} (Oxford University Press, Oxford, 2009).

\bibitem{Mello} E. R. Bezerra de Mello, V. B. Bezerra, A. A. Saharian and A.
S. Tarloyan, Phys. Rev. D \textbf{74}, 025017 (2006).

\bibitem{Brev95} I. Brevik and T. Toverud, Class. Quantum Grav. \textbf{12},
1229 (1995).

\bibitem{Mello1} E. R. Bezerra de Mello, V. B. Bezerra and A. A. Saharian,
Phys. Lett. B \textbf{645}, 245 (2007).

\bibitem{Aram1} E. R. Bezerra de Mello, V. B. Bezerra, A. A. Saharian and A.
S. Tarloyan, Phys. Rev. D \textbf{78}, 105007 (2008).

\bibitem{Beze10} E. R. Bezerra de Mello, V. B. Bezerra, A. A. Saharian and
V. M. Bardeghyan, Phys. Rev. D \textbf{82}, 085033 (2010); S. Bellucci, E.
R. Bezerra de Mello and A. A. Saharian, arXiv:1101.4130, to appear in Phys.
Rev. D.

\bibitem{Mello3} E. R. Bezerra de Mello and A. A. Saharian, Phys. Lett. B
\textbf{642}, 129 (2006).

\bibitem{Fucc11} G. Fucci and K. Kirsten, arXiv:1101.5409.

\bibitem{Grad} I. S. Gradshteyn and I. M. Ryzhik, \textit{Table of
Integrals, Series and Products} (Academic Press, New York, 1980).

\bibitem{Pru} A. P. Prudnikov, Yu. A. Brychkov and O. I. Marichev, \textit{%
Integrals and Series: Special Functions}. Vol 2, (Moscow, Nauka, 1983) (in
Russian).

\bibitem{Jean} J. Spinelly and E. R. Bezerra de Mello, J. High Energy Phys.
\textbf{09}, 005 (2008).

\bibitem{Abra} M. Abramowitz and I. A. Stegun, \textit{Handbook of
Mathematical Functions} (Dover, New York, 1972).

\bibitem{SahaRev} A. A. Saharian, \textit{The Generalized Abel-Plana Formula
with Applications to Bessel Functions and Casimir Effect} (Yerevan State
University Publishing House, Yerevan, 2008); Preprint ICTP/2007/082;
arXiv:0708.1187.

\bibitem{Cristine} C. A. B. Bayona, C. N. Ferreira and V. J. V. Otoya,
Class. Quantum Grav. \textbf{28}, 015011 (2011).
\end{thebibliography}
\end{document}